\documentclass[aps,floatfix,pra,twocolumn,nofootinbib]{revtex4}%floatfix

\usepackage{graphicx}
\usepackage{graphics}
\usepackage{amssymb}
\usepackage{amsmath}
\usepackage{epsfig}
\usepackage{latexsym}

\newcommand{\ket}[1]{|#1\rangle}
\newcommand{\bra}[1]{\langle #1|}

\newcommand{\op}[1]{\hat{#1}}

\newcommand{\jxo}{\hat{J}_{x1}}

\newcommand{\jzo}{\hat{J}_{z1}}

\newcommand{\jz}{\hat{J}_{z}}
\newcommand{\jxt}{\hat{J}_{x2}}

\newcommand{\jzt}{\hat{J}_{z2}}

\newcommand{\lxo}{L_{x1}}
\newcommand{\lyo}{L_{y1}}
\newcommand{\lzo}{L_{z1}}
\newcommand{\lxt}{L_{x2}}
\newcommand{\lyt}{L_{y2}}
\newcommand{\lzt}{L_{z2}}

\newcommand{\eqrf}[1]{(\ref{#1})}

\newcommand{\bea}{\begin{eqnarray}}
\newcommand{\eea}{\end{eqnarray}}
\newcommand{\ra}{\rightarrow}
\newcommand{\la}{\leftarrow}
\newcommand{\ua}{\uparrow}
\newcommand{\da}{\downarrow}
\newcommand{\ox}{\op{X}}
\newcommand{\oy}{\op{Y}}

\renewcommand{\[}{\begin{equation}}
\renewcommand{\]}{\end{equation}}

\newtheorem{con}{Conjecture}

\begin{document}

\title{Quantum entanglement and fixed-point bifurcations}
\author{Andrew P. Hines}
\email{hines@physics.uq.edu.au}

\author{Ross H. McKenzie}
\author{G.J. Milburn}
\affiliation{Centre for Quantum Computer Technology, School of Physical Sciences, The University of Queensland, St Lucia, QLD 4072, Australia}
\date{\today}

\begin{abstract}
 How does the classical phase space structure for a composite system relate to the entanglement characteristics of the corresponding quantum system? We demonstrate how the entanglement in nonlinear bipartite systems can be associated with a fixed point bifurcation in the classical dynamics. Using the example of coupled giant spins we show that when a fixed point undergoes a supercritical pitchfork bifurcation, the corresponding quantum state - the ground state -  achieves its maximum amount of entanglement near the critical point. We conjecture that this will be a generic feature of systems whose classical limit exhibits such a bifurcation.
\end{abstract}

\maketitle

\section{Introduction}

With the advent of quantum information theory, entanglement is now regarded as a physical resource that can be utilized to perform numerous quantum computational and communication tasks \cite{Nie00}. This has in turn led to the study of the entanglement characteristics of various systems, and in turn, how these characteristics relate to more fully understood properties of the system. Such studies are two-fold beneficial - further elucidating the nature of entanglement as well as providing a new approach to the study of complex, quantum many-body systems.

One area where such an approach has had some success is in the study of {\it quantum phase transitions} (QPTs) - qualitative changes in the ground state of a multi-partite system induced by the variation of some external parameter \cite{Sac99}. There have been many recent studies relating entanglement and QPTs (see \cite{ON02,OAF02,VLR02,LRV03,CM03,Hin03,DFJ+02}).

Generally it has been found that in infinite systems that undergo a quantum phase transition at a critical parameter value, $\lambda=\lambda_c$, the entanglement as a function of $\lambda$ is a maximum at $\lambda_c$. Several examples include:
(i) The single site entanglement and the next nearest neighbor concurrence of the transverse Ising chain \cite{ON02,OAF02,VLR02} (although the nearest neighbor concurrence does not have its maximum value at $\lambda=\lambda_c$, its first derivative with respect to $\lambda$ does \cite{OAF02}), (ii) the entropy of entanglement of half of a XXZ spin chain in a magnetic field \cite{LRV03}, and (iii) the entropy of entanglement of a single qubit with a bath of oscillators (the spin-boson model) \cite{CM03}.
Such systems demonstrate a correspondence between quantum critical phenomena and entanglement.

A QPT corresponds to a qualitative change in the ground state as a system parameter is varied. In the classical regime, minimum energy coordinates correspond to elliptic (stable) fixed points. As a parameter in the system is varied, fixed points may undergo {\it bifurcation} \cite{Gle94} - a loss of stability, the emergence of new fixed points - at some critical value of the parameter. This corresponds to a qualitative change in the phase space structure of the system.

In this article we consider the ground state of a system whose classical limit exhibits a bifurcation where a single elliptic fixed point loses its stability while two new, elliptic points emerge -- a so-called supercritical pitchfork bifurcation. Elliptic fixed points can be associated with the ground state of the quantized system. Subsequently we  expect to see some signature of the classical bifurcation in the quantum ground state, around the critical point. We argue that this signature is a peak in the entanglement with respect to the bifurcation parameter.

Schneider and Milburn alluded to such a correspondence in their work on the Dicke model \cite{Sch01}. It was shown that the entanglement in the steady state of this system is a maximum for the parameter value corresponding to a bifurcation of the fixed points in the corresponding classical dynamics. It was conjectured that the loss of stability of a classical fixed point due to such a bifurcation will generically be associated with entanglement in the steady state of the full quantum system. We show that it is specifically the pitchfork nature of this bifurcation that is responsible for the peak in the ground state entanglement.

\begin{figure}[h!]
\begin{center}
\scalebox{.4}{\includegraphics{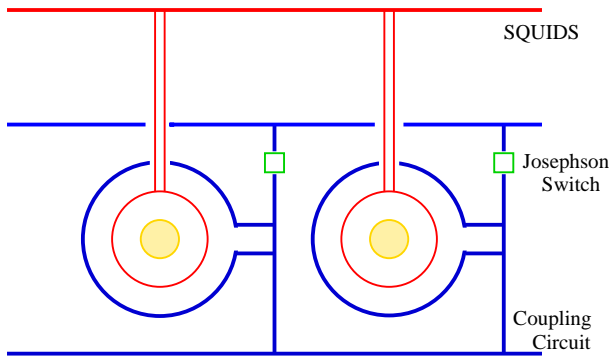}}
\end{center}
\caption{(Color online) A schematic diagram of the coupled qubit realization of Ref.\cite{Tej00}. The magnetic clusters (the qubits) are coupled to superconducting loops of micro-SQUIDS and arranged in a 1D lattice. Josephson junction switches are used in the coupling circuits, as shown.}
\label{cir}
\end{figure}

To demonstrate this, we use the example of coupled giant spins. This system is motivated by a proposed physical implementation for quantum computation \cite{Tej00}. In this proposal, qubits are realized by magnetic clusters - nanometer scale molecular clusters that have all the attributes of mesoscopic systems such as angular momentum and magnetic moment. Two qubit gates are constructed by the coupling of the clusters via \emph{superconducting quantum interference devices} (or SQUIDs), as shown in figure \ref{cir}.

 The strength of the coupling is dependent upon the super-current induced in the loop by one spin and the field this produces at the other site \cite{Tej00}. This field induces a rotation about the `$z$'-axis of one cluster with frequency proportional to the $z$-component of the angular momentum of the other cluster. In this way, we can imagine the system as a set of `spinning tops', coupled via a nonlinear interaction, described by $\op{H}_{int}=\jz\otimes\jz$.

The article is organized as follows: In Sec. \ref{Sec::cpt} we introduce the coupled tops model. We begin with a review of the classical dynamics based on Skellett and Holmes \cite{axe02}, focussing on the bifurcation analysis. This is followed by a calculation of the ground state entanglement for the quantum system, as the bifurcation parameter is varied. Here the link between the ground state entanglement and the bifurcation is first established. Using the Husimi distribution (or $Q$-function) to represent the ground state in phase space, we demonstrate how the structure of the ground state is related to the classical fixed points. The aim of Sec. \ref{Sec::QFE} is to explain the observed behaviour and argue its generality. We finish with a discussion of our results and future directions.

\section{The Coupled Tops Model}\label{Sec::cpt}

The simplest Hamiltonian describing the coupled tops is a generalization of the $N = 2$ case of the transverse field quantum Ising model, i.e.
\begin{equation}
\op{H} = \omega \hat{J}_{x} \otimes \op{I} + \omega \op{I}\otimes  \hat{J}_{x} + \frac{\chi}{j}  \hat{J}_{z}\otimes  \hat{J}_{z}
\label{ham}
\end{equation}
where the angular momentum operators $\op{J}_{a}$ satisfy the $SU(2)$ commutation relations
$ [ \hat{J}_{x}, \hat{J}_{y}] = i \hat{J}_{z}$ (and cyclic permutations). Setting $\mu = \frac{\chi}{\omega}$, results in a one-parameter Hamiltonian
\begin{equation}\label{ham2}
\op{H} = \jxo + \jxt + \frac{\mu}{j}\jzo\jzt
\end{equation}
where we make use of the notation $\hat{J}_{a1} =\hat{J}_{a} \otimes \op{I}$ and $\hat{J}_{a 2} = \op{I} \otimes \hat{J}_{a}$, $a = x,y,z$, such that the subscript $1$ ($2$) refers to subsystem $1$ ($2$).

The square of the total angular momentum of the individual tops,
\begin{equation}
\op{J}_i^{2} = \op{J}_{xi}^2+\op{J}_{yi}^2+\op{J}_{zi}^2
\end{equation}
satisfy
\[
\left[ \op{J}_{1}^{2},\op{H} \right] = \left[ \op{J}_{2}^{2},\op{H} \right] =0.
\]
so are constants of the motion. Here the tops are identical, such that $\hat{J}_1^2 = \hat{J}_2^2 = j(j+1)$. This allows the system to be represented in the basis of tensor products of the $\op{J}_z$ eigenstates, $|j,m\rangle \otimes |j,n\rangle \equiv |m,n\rangle$, where $-j\leq m,n\leq j$. Note that the coupling term in the Hamiltonian is scaled with $j$, to allow the classical limit to be taken.For $j = \frac{1}{2}$, the Hamiltonian (\ref{ham}) is analogous to the quantum Ising model for $2$ spins, studied in Ref. \cite{Gun01}.

Interestingly, the square of the \emph{total} angular momentum of the system,
\begin{equation}
\op{J}^{2} = \op{J}_{1}^{2} + \op{J}_{2}^{2} + 2\mathbf{\op{J}_{1}} \cdot \mathbf{\op{J}_{2}},
\end{equation}
is not a constant of the motion. The reason is that an external control is required to couple the tops. In our motivating example this is the circuit that inductively couples the tops. This is similar to the situation for two interacting qubits. A general two qubit gate, such as a controlled-NOT gate does not conserve total angular momentum either (though the gate can fix the singlet and triplet subspaces). Again this is due to external interactions that control the gate.

The classical analogue of this system has been rigorously studied by Skellett and Holmes \cite{axe02}. We now derive the semiclassical limit of Hamiltonian (\ref{ham2}), showing its correspondence to the model of Ref. \cite{axe02}, and review the key points of this analysis relevant to our work.

\subsection{Classical Description}

The semiclassical limit of the coupled tops system corresponds to the limit of $j\rightarrow \infty$. To obtain the semiclassical model, we express the classical coordinates as $L_{a\alpha} = \langle \hat{J}_{a\alpha}\rangle /j$. In the limit, this allows the factorization of all moments, i.e. $\langle\hat{J}_{x1}\hat{J}_{z1}\rangle/j^2 \rightarrow L_{x1}L_{z1}$ (for details see Appendix A). The classical equations of motion are obtained from the Heisenberg operator equations of motion by taking expectation values and applying the factorization rule above. The corresponding semiclassical Hamiltonian is
\[
E = L_{x1} + L_{x2} + \mu L_{z1}L_{z2}
\]
where $E = \langle \hat{H}\rangle/j$ with the spherical constraint, $L_{xi}^{2} + L_{yi}^{2} + L_{zi}^{2} =1$.

  The analysis of the corresponding classical system \cite{axe02} has shown that the non-linearity of the interaction term leads to chaotic motion for given parameter ranges and initial conditions. More relevant in this context is the existence of a supercritical pitchfork bifurcation, at a critical value of the coupling parameter. From Ref.\cite{axe02} for the semiclassical system, the critical value is $\mu_{c} = 1$. Below this critical value the dynamics of the system is predominantly regular while above the phase space is mixed, with extensive regions of chaotic motion.
\begin{figure}[t]
\begin{center}
\scalebox{0.55}{\includegraphics{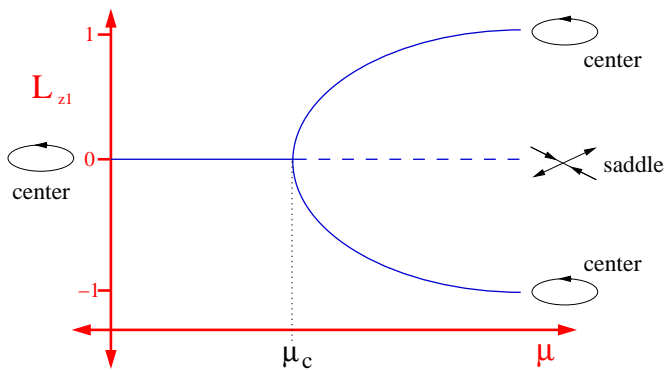}}
\end{center}
\caption{(Color online) The pitchfork bifurcation at $\mu_{c}=1$ for $\lzo$. The elliptic (stable) fixed points are centres, indicating that close to the fixed point, the motion is periodic, while the hyperbolic (unstable) point is a saddle. This diagram is the same for both values of $L_{x1}=L_{x2} = \pm 1$.}
\label{bifur}
\end{figure}

The fixed points of the system are found by setting the equations of motion to zero, to determine the coordinates where the phase space flow is zero. From Skellett and Holmes \cite{axe02}, there are four solutions which exist for all values of the coupling parameter $\mu$, given by
\begin{equation}\label{fp1}
\lxo = \pm 1 , \lxt = \pm 1, \lzo=\lyo=\lzt=\lyt = 0.
\end{equation}
At the critical value, $\mu_{c} = 1$  the two fixed points at $\lxo = \lxt = 1$ and $\lxo = \lxt = -1$ bifurcate, resulting in the emergence of a further {\it four} fixed points, located at
\begin{eqnarray}
\lxo=\lxt=\frac{1}{\mu},& \lzo=\lzt=&\pm\sqrt{1-\frac{1}{\mu^{2}}},\nonumber\\
 &\lyo=\lyt=0,\label{AB}& \\
\lxo=\lxt=-\frac{1}{\mu},& \lzo=-\lzt=&\pm\sqrt{1-\frac{1}{\mu^{2}}},\nonumber\\
& \lyo=\lyt=,0\label{CD}&
\end{eqnarray}
which exist for all $\mu > 1$. The stability of the fixed points is determined by analysis of the eigenvalues of the linearized matrix about each fixed point \cite{Gle94}.

In Ref. \cite{axe02}, it was shown that the two fixed points \eqrf{fp1} with $\lxo=-\lxt$ are {\it unstable} for all values of $\mu$. The points, $\lxo=\lxt = \pm 1$ are stable for $\mu < 1$, becoming unstable at $\mu_c$. The emergent fixed points are all stable. This implies that the bifurcations occurring at $\lxo=\lxt = \pm 1$ are supercritical pitchfork bifurcations, as illustrated in figure \ref{bifur}.

Since the total angular momenta of the two tops are conserved, their dynamics of each are constrained to the unit sphere, in angular momentum space. It is possible to reformulate the dynamics in terms of spherical polar coordinates - the polar angle from the positive $L_{zi}$ axis, $0\leq \theta_{i} \leq \pi$, and the azimuthal angle in the $L_{xi}-L_{yi}$ plane (from the positive $L_{xi}$ axis), $0\leq \phi_{i} \leq 2\pi$. These coordinates give the angular momentum components via
\bea
L_{xi} &=& \sin \theta_{i}\cos \phi_{i} \nonumber\\
L_{yi} &=& \sin \theta_{i}\sin \phi_{i}\\
L_{zi} &=& \cos \theta_{i}\nonumber.
\eea

For all fixed points $\lyo=\lyt=0$, which in spherical polar coordinates, corresponds to $\phi_{1},\phi_{2}=0$ or $\pi$. Thus, we can view the fixed points as lying on the unit circle in the $L_{xi}-L_{zi}$ planes, characterized by the polar angles, $\theta_{i}$. For $\mu$ below the critical coupling there are two, stable fixed points, both of which lie at the `equator' of these unit circles ($\theta_{i}=\frac{\pi}{2}$) at $\lxo=\lxt=1$ ($\phi_{1}=\phi_{2}=0$) and at $\lxo=\lxt=-1$ ($\phi_{1}=\phi_{2}=\pi$). Following the notation used by Skellett and Holmes \cite{axe02}, we denote the two states by ($\ra\ra$) and ($\la\la$) respectively, corresponding to the direction of the angular momentum vector.

\begin{figure}[t]
\begin{center}
\scalebox{0.4}{\includegraphics{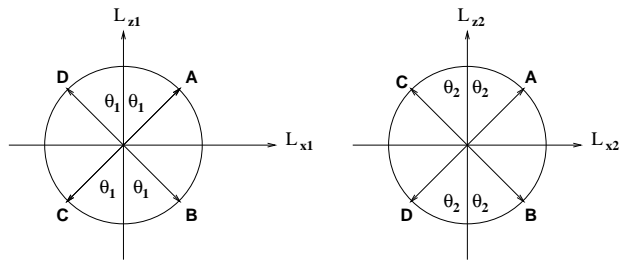}}
\end{center}
\caption{The four stable fixed points for $\mu>1$. As $\mu\rightarrow \infty$, $\theta_{1}$,$\theta_{2}\rightarrow 0$, resulting in fixed points with angular momentum solely in the $L_{z}$ directions, denoted $(\ua\ua),(\ua\da),(\da\ua)$ and $(\da\da)$.}
\label{fps}
\end{figure}

For $\mu$ greater than the critical value, there are four stable fixed points, whose positions are shown in figure \ref{fps}.
The points labeled $A$ and $B$ correspond to Eq. \eqrf{AB} and points $C$ and $D$ to Eq. \eqrf{CD}. Clearly as $\mu \ra \infty$, $\theta_{i}\ra 0$ (in figure \ref{fps}) and, in the pictorial (arrow) notation of above, we have the four fixed points at the `poles' of the spheres, in the four combinations $(\ua\ua),(\ua\da),(\da\ua)$ and $(\da\da)$, all of which are stable.

The semiclassical analysis we have presented here is just a brief summary of those aspects most relevant for this paper. For an in-depth analysis of the classical dynamics of this system we again refer the reader to Skellett and Holmes \cite{axe02}.

The bifurcating fixed point at $L_{x1}=L_{x2} = -1$ (for $\mu < \mu_c$) is the minimum energy point and so corresponds to the quantum ground state. We now consider the quantum regime, and the entanglement between the spins as a function of the coupling strength, $\mu$.

 \subsection{Ground State Entanglement}

The ground state is computed by direct numerical diagonalization of the Hamiltonian (\ref{ham2}). The entanglement measure we employ is the entanglement of formation, which, for pure bipartite states, is equivalent to the entropy of entanglement \cite{pres,Nie00}
\[
S(\rho_i) = -\textrm{Tr}\left(\rho_i \log \rho_i \right)
\]
where $\rho_i = \textrm{Tr}_i\left(\rho\right)$ is the reduced density operator and the $\log$ is to base $2$.

The case of $j = 1/2$, i.e. a two-site transverse field Ising model, was considered by Gunlycke {\it et al.} \cite{Gun01}. It was shown in this case that the ground state entanglement is zero for zero coupling ($\mu = 0$) then increases as the coupling increases, to asymptote to the maximal value of $1$ as $\mu \ra \infty$.

\begin{figure}[t]
\begin{center}
\scalebox{0.48}{\includegraphics{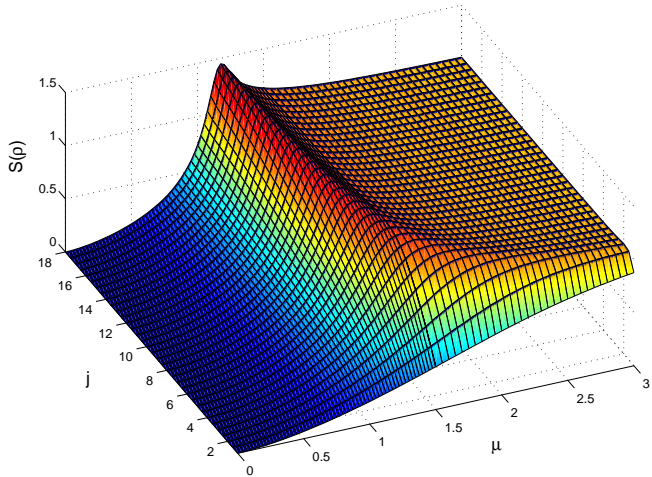}}
\end{center}
\caption{(Color online) Variation in the entropy of entanglement of the ground state for with respect to the coupling strength $\mu$ and the total subsystem angular momentum, $j$. Note that as the system becomes more classical as $j$ increases that the peak in the entanglement versus $\mu$ becomes more evident.}
\label{fig::entvarj}
\end{figure}

For $j>1/2$,  the entanglement of the ground state with respect to the interaction strength, $\mu$ takes on a new characteristic (see figure \ref{fig::entvarj}).

For $\mu = 0$, the ground state is simply a tensor product of the minimal $\op{J}_{xi}$ weight states, $|-j,-j\rangle_x$, which is separable.  As $\mu\ra\infty$, the ground state approaches the superposition, $\left(|j,-j\rangle + |-j,j\rangle\right)/\sqrt{2}$, so the entanglement still asymptotes to $1$. However for $j>1/2$ the entanglement now peaks at a finite value of $\mu$. The height of this peak grows with the value of $j$ and will approach infinity in the limiting case.

\begin{figure}[t!]
\begin{center}
\scalebox{0.65}{\includegraphics{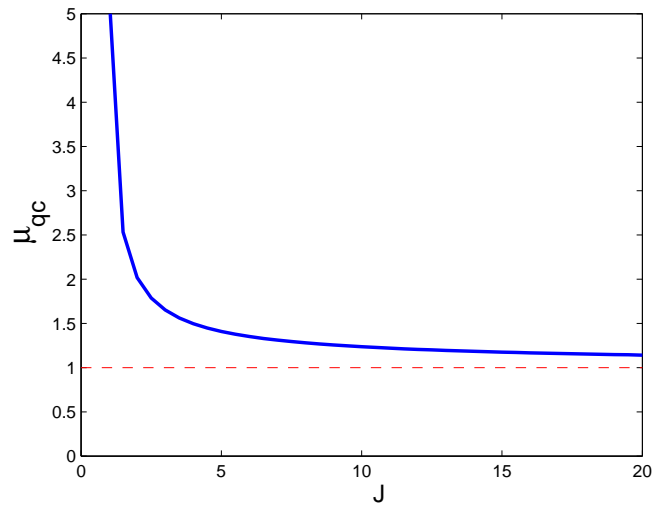}}
\end{center}
\caption{(Color online) The variation in the quantum critical parameter, $\mu_{qc}$, the value of $\mu$ for maximum ground state entanglement, with total angular momentum $j$. In the classical limit, $j\rightarrow\infty$, $\mu_{qc}$ approaches the classical critical parameter, where the bifurcation occurs.}
\label{fig::muqc}
\end{figure}

We let $\mu_{qc}$ denote the quantum critical parameter, defined as the coupling value at which the ground state entanglement is maximum. Figure \ref{fig::muqc} shows the limiting behaviour of $\mu_{qc}$. For larger $j$, the maximum entanglement occurs near the bifurcation point and in the limit $j\rightarrow\infty$, the quantum critical point approaches the classical bifurcation point,
\[
\mu_{qc} \rightarrow \mu_c.
\]
In the semiclassical limit, the quantum critical point corresponding to maximum ground state entanglement is the classical bifurcation point.
To understand how the fixed point structure manifests in the quantum regime, we need to consider the structure of the quantum state in phase space.

\subsection{Ground State in Phase Space}

\begin{figure*}[t!]
%\begin{flushleft}
\begin{center}
\scalebox{0.38}{\includegraphics{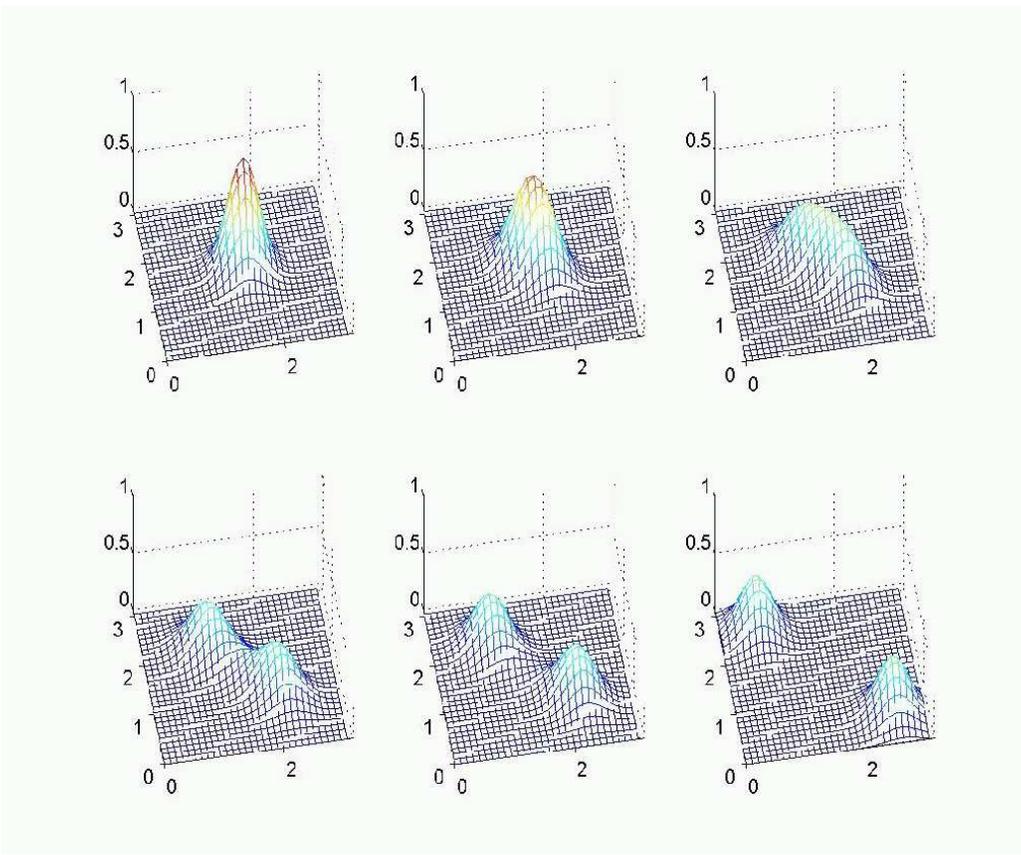}}
%\end{flushleft}
\end{center}
\caption{(Color online) The $\phi_{1}=\phi_{2}=\pi$ cross-sections of the ground state $Q$-function for a coupling strength $\mu$ equal to (a) $0$, (b) $0.7$, (c) $1$, (d) $1.184$, (e) $1.25$ and (f) $1.55$, for $j = 14$. The $x$ and $y$ axes correspond to $\theta_1$ and $\theta_2$ respectively. The entanglement of formation is $0$ for (a), approximately $1$ for (f), and maximum for (d), when the distribution is the most delocalized. We see that as the fixed point bifurcates, so does the ground state $Q$-distribution, from the single-peak to the twin-peaked structure.}
\label{fig::evoQ}
\end{figure*}

Fundamental to any comparison of classical and quantum dynamics is some notion of the quantum analogue of a classical joint phase-space probability distribution. We choose the Husimi or $Q$-function \cite{Hus40} as the appropriate quantum analogue of the classical phase-space density, following ref.'s \cite{Lat93,Mil86,San89,Lin90,Lee93,MLSK01}.

For systems described in spherical phase-space, Appleby \cite{App00} demonstrated that the \emph{positive operator valued measurement} (POVM) for optimal simultaneous measurements of angular momentum components is given by $\op{E}(z) = \ket{z}\bra{z}$, where $\ket{z}$ are the $SU(2)$, coherent states \cite{Per86},
\begin{equation}
|z\rangle = (1+|z|^2)^{-j}e^{z\hat{J}_+}|j,-j\rangle \label{suco}
\end{equation}
with $\ket{j,-j}$ the lowest weight eigenstate of $\jz$ and $z$, the stereographic projection of the sphere onto the plane,
\[
z(\theta,\phi)=e^{-i\phi} \tan\frac{\theta}{2}.
\]
The probability distribution for measurements defined by this POVM is the angular momentum representation $Q$-function, which for pure state $\rho = \ket{\psi}\bra{\psi}$ is
\[\label{Qf}
Q_{\psi}(z)=Tr\left(\rho\op{E}(z)\right)=\left|\langle z\ket{\psi}\right|^{2}.
\]

For our bipartite system we define the two-body $Q$-function in terms of the coherent states
\[\label{many_cos}
\ket{\boldsymbol{z}} = \ket{z_1}\otimes\ket{z_2}.
\]
Since the $Q$-function for the coupled tops is in $4$ phase-space dimensions, we may only display cross-sections graphically. We have calculated the $\phi_{1}=\phi_{2}=\pi$ cross-section of the ground state $Q$-function for various values of the coupling strength, $\mu$. This cross-section was chosen since this is the plane in which the bifurcating fixed points are located.

As evidenced in figure \ref{fig::evoQ}, the structure of the ground state in phase space is intimately related to the fixed point structure and bifurcation. For small $\mu$, the $Q$-distribution of the ground state is localized around the fixed point at $\theta_1 = \theta_2 = \pi/2$ which is elliptic for $\mu < \mu_c$. For $\mu$ much larger, beyond the bifurcation, the $Q$-distribution is twin-peaked, localized around the two emergent elliptic fixed points. Put simply, in these two extremes, the distribution is localized around the elliptic fixed points. In between, the distribution is spread between the three fixed points and it is this region in which the degree of entanglement is greatest. In fact, we will argue that the entanglement is maximum when the distribution is at its most delocalized \cite{Sug01}.

Under this assumption, the maximum ground state entanglement will occur when $\mu$ is above, but close to the critical point. This is when the $Q$-distribution is at its most delocalized, smeared between the three fixed points. As $j$ becomes very large, and the Heisenberg uncertainty limit allows for a higher degree of localization of the distributions, this quantum critical point approaches the classical bifurcation point. In this semiclassical regime, the greatest delocalization occurs closer and closer to the bifurcation, as $j\ra\infty$.

For this simple coupled tops model we have demonstrated that the entanglement characteristics of the ground state can be associated with the supercritical pitchfork bifurcation. To validate and generalize this result, we need to consider why the ground state $Q$-distribution is related to the fixed point structure, and how this structure corresponds to the degree of entanglement.

\section{The Ground state $Q$-function and Entanglement}\label{Sec::QFE}

There has been considerable effort in relating quantum eigenstates to classical phase space structures (for example see \cite{BTU93,Ber77,KHSW00}). The \emph{Einstein-Brillouin-Keller} (EBK) approximation \cite{Haa91} provides for a semiclassical quantization of classically integrable systems, whereby eigenstates are identified with closed loops around invariant tori in phase space. For non-integrable systems, the Gutzwiller trace formula allows for semiclassical approximations of energy spectra, but not the structure of the corresponding eigenstates \cite{Haa91}.

For systems exhibiting mixed phase space, while no general quantization procedure exists, there is an understanding that eigenstates can be separated into regular and irregular/chaotic \cite{Per73} (and in-between or hierarchical \cite{KHSW00}) groups. Regular states are supported by classical tori obeying EBK quantization, whereas chaotic states are associated with chaotic phase space regions. In the Husimi representation, regular states are seen as localized on regular trajectories, whereas chaotic states are somewhat evenly distributed over the chaotic region of phase space \cite{Shn74}.

Classically, the minimum energy trajectories correspond to elliptic (stable) fixed points. Thus, it would seem reasonable to infer that the quantum ground state would be localized around such points.

 The aim of this section is to demonstrate that for Hamiltonians whose classical analogue exhibits (minimal energy) elliptic fixed points, the $Q$-function of the ground state is peaked on the phase space coordinates corresponding to such fixed points. This statement has non-trivial consequences when the classical Hamiltonian displays a \emph{bifurcation} of fixed points, which, in the quantum regime, will correspond to a qualitative change in the structure of the ground state phase-space distribution. Subsequently, we relate the structure of the $Q$-function to  entanglement properties.

\subsection{Ground state $Q$-function and fixed points}

In reformulating quantum mechanics in phase space, the coherent states provide a natural phase space structure for a given quantum system, as well as useful distributions based in the coherent state representation \cite{ZFG90}, two of which, we briefly discuss now.

An operator $\op{O}$ can be expressed in the diagonal form
\begin{equation}
\op{O} = \int O_P(z) |z\rangle\langle z| d\mu(z)
\end{equation}
called the $P$ representation, where $|z\rangle$ (in complex variable $z$) are the coherent states and $d\mu(z)$ is the corresponding measure. The density operator is given by
\begin{equation}
\rho = \int P(z) |z\rangle\langle z| d\mu(z).
\end{equation}
The $Q$-representation of an operator is defined as
\begin{equation}
\op{O} \rightarrow O_Q(z) = \langle{z}|\op{O}|z\rangle
\end{equation}
and the density operator in the representation is denoted by $Q(z)$.
The statistical average of an operator in the two representations is given by
\begin{eqnarray*}
\langle\op{O}\rangle = \textrm{Tr}(\rho\op{O}) & = &\int Q(z)O_P(z)d\mu(z)\\
& = & \int P(z)O_Q(z)d\mu(z).
\end{eqnarray*}

To determine the ground state of a given Hamiltonian, $\op{H}$, in this phase space formulation we seek the distribution that minimizes the average energy, i.e. determine the distribution $Q(z)$ that minimizes
\begin{equation}
\langle\op{H}\rangle = \int Q(z)H_P(z)d\mu(z)
\end{equation}
for a given $H_P(z)$.

Upper and lower bounds on the ground state energy, $E_0$, are determined by minimizing the $Q$ and $P$ representations of the Hamiltonian \cite{ZFG90},
\begin{equation}
\min_{z} H_Q(z) \geq E_0 \geq \min_{z} H_P(z).
\end{equation}

In the semi-classical limit (i.e $\hbar \rightarrow 0$ for Heisenberg-Weyl, $J \rightarrow \infty$ for $SU(2)$ coherent states) of the appropriately scaled Hamiltonian, both the $Q$ and $P$ representations converge to the classical Hamiltonian, $\mathcal{H}(z)$. So the problem of finding the ground state in the phase space representation as we approach the semiclassical limit reduces to finding the distribution $Q(z)$, minimizing the functional,
\begin{equation}
\mathcal{F}[Q(z)] = \int d\mu(z)~Q(z)\mathcal{H}(z)\label{eq:HQ_min}
\end{equation}
where $\mathcal{H}(z)$ is the semiclassical Hamiltonian. This distribution must satisfy
\begin{equation}
\int d\mu(z)~Q(z) = 1,\qquad 0 \leq Q(z) \leq 1
\end{equation}
as well as the uncertainty principle, which excludes the possibility of delta functions.

Let us consider a suitably scaled Hamiltonian with semiclassical limit, $\mathcal{H}(z)$, which exhibits a global minimum at $z = z_0$. This requires that
\begin{equation}
\left. \frac{\partial \mathcal{H}}{\partial z}\right|_{z_0} = \left. \frac{\partial \mathcal{H}}{\partial \bar{z}}\right|_{z_0} = 0
\end{equation}
and
\begin{equation}
\left(\left. \frac{\partial^2 \mathcal{H}}{\partial z\partial \bar{z}}\right|_{z_0}\right)^2 - \left. \frac{\partial^2 \mathcal{H}}{\partial z^2}\right|_{z_0}\left. \frac{\partial^2 \mathcal{H}}{(\partial \bar{z})^2}\right|_{z_0} > 0.
\end{equation}
This corresponds to an elliptic fixed point in the phase space portrait of the classical dynamics.
Taylor expanding $\mathcal{H}(z)$ about the fixed point gives
\begin{eqnarray*}
\mathcal{H}(z,\bar{z}) &=& \mathcal{H}(z_0,z_0^*) + |z - z_0|^2 \left. \frac{\partial^2 \mathcal{H}}{\partial z\partial \bar{z}}\right|_{z_0} \\& & +\frac{1}{2}(z - z_0)^2 \left. \frac{\partial^2 \mathcal{H}}{\partial z^2}\right|_{z_0}\\ & & + \frac{1}{2}(\bar{z} - \bar{z}_0)^2\left. \frac{\partial^2 \mathcal{H}}{(\partial \bar{z})^2}\right|_{z_0} + \textrm{h.o.t.}
\end{eqnarray*}
which can be written as
\begin{equation}\label{eq::expansion}
\mathcal{H}(z,\bar{z}) \approx \mathcal{H}(z_0,z_0^*) + \left[\begin{array}{cc} z-z_0 & \bar{z}-z_0^* \end{array}\right] \mathbb{D} \left[\begin{array}{c} z-z_0 \\ \bar{z}-z_0^* \end{array}\right]\nonumber,
\end{equation}
where
\begin{equation}
\mathbb{D} = \left[ \begin{array}{cc}
\left. \frac{\partial^2 \mathcal{H}}{\partial z^2}\right|_{z_0} & \left. \frac{\partial^2 \mathcal{H}}{\partial z\partial \bar{z}}\right|_{z_0}\\
\left. \frac{\partial^2 \mathcal{H}}{\partial z\partial \bar{z}}\right|_{z_0} & \left. \frac{\partial^2 \mathcal{H}}{(\partial \bar{z})^2}\right|_{z_0}
\end{array}\right], \end{equation}
such that $\textrm{det}(\mathbb{D}) < 0$. In other words, to second order around the elliptic fixed point, $\mathcal{H}(z,\bar{z})$ is parabolic, with minimum at $z_0$.

Now consider the minimization problem of Eq. (\ref{eq:HQ_min}). If the system under investigation has only a solitary global minima, since $Q(z)$ takes only positive values, the distribution that minimizes Eq. (\ref{eq:HQ_min}) will be as sharply peaked as possible on the coordinates of the minima. The ground state $Q$-function will be peaked on the coordinates of the fixed point.

What if there are more than one (degenerate) global minima? In this case, consider expansions around each of the fixed points $z_k$ following (\ref{eq::expansion}). We are interested in the ground state in a semiclassical regime where the $Q$-functions can be highly localized. We thus approximate $\mathcal{H}$ as a piecewise sum of the expansions around each of the degenerate fixed points. The functional (\ref{eq:HQ_min}) is thus written as the summation
\begin{equation}
\mathcal{F}[Q(z)] = \sum_k \int_{\mathcal{R}_k}d\mu(z)~Q_k(z)\mathcal{H}_k(z)
\end{equation}
where $\mathcal{H}_k(z)$ is the expansion around the fixed point $z_k$ and $Q_k(z)$ is the distribution defined in the phase-space region $\mathcal{R}_k$. Minimizing each term in this summation gives  in turn a piecewise defined distribution, sharply peaked around the coordinates of each of the fixed points.

 The above statements are by no means rigorous. We are considering the problem in a semiclassical regime, which allows us to take simple approximations for the Hamiltonian, and simplifies the structure of the corresponding ground state $Q$-functions. However, as we take the semiclassical limit ($\hbar \rightarrow 0$, $j\rightarrow\infty$), the $Q$-function will merge \emph{smoothly} from a true quantum ground state distribution to the more semiclassical structure described above.

As we move in the other direction, from the semiclassical to the quantum, we expect to observe remnants of the semiclassical structure. The $Q$-function, while perhaps not sharply peaked, will be concentrated around the fixed point coordinates.

These observations lead us to make the following conjecture:

\begin{con}[Fixed point correspondence]\label{t1}

Let $\op{H}$ be a Hamiltonian whose classical analogue is defined as $\mathcal{H}(\chi,\chi^*)=\bra{\chi}\op{H}\ket{\chi}$, where $\left\{\ket{\chi}\right\}$, is the set of coherent states corresponding to the topology of the phase-space of $\mathcal{H}(\chi,\chi^*)$. Then the Husimi distribution of the ground state of $\op{H}$ will be concentrated around the phase-space coordinates corresponding to the fixed point(s) of $\mathcal{H}$.
\end{con}
While we have not provided a rigorous proof of this statement, we have argued above that such a notion is indeed plausible.

This conjecture by no means provides detailed information about the ground state, but instead provides a guideline for the phase-space structure of the ground state. As in Ref. \cite{ZFG90}, where the use of the $Q$-representation of the Hamiltonian as a guide to studying ground state phase transitions was advocated, this guideline become most useful when the fixed points vary according to some critical parameter i.e. \emph{bifurcation}. This can provide for a variational approach to approximating the ground state via coherent states (see Ref. \cite{ZFG90}).

For our purposes, this link between fixed points and the phase-space structure can provide a qualitative understanding of the effect of the bifurcation on the ground state entanglement. The connection to entanglement is made via the work of Sugita \cite{Sug01}, which we now discuss.

\subsection{Entanglement and the $Q$-function}

When dealing with $Q$-functions for composite systems, there are several ways to generalize the idea of coherent states to many-body systems. In the next section, we follow Sugita \cite{Sug01} by constructing coherent states based on the single-particle transformation group. This means the coherent states are independent-particle states, and hence separable, allowing the entanglement to be related to the structure of the $Q$-function.

Sugita \cite{Sug01} constructs the many-body coherent states based on the single-particle transformation group, following the group-theoretical construction of Perelomov \cite{Per86}. As an illustrative example, consider a system of $n$ qubits (basis states $|\ua\rangle,|\da\rangle$). The local unitary (or single-particle) transformation group is $SU(2)^{\otimes n}$. Coherent states are generated by applying this group to the lowest weight state $|\da\rangle^{\otimes n}$. In this way, we generate all (pure) separable states. From this definition, coherent states are thus equivalent to separable states. We note that our arguments in the above section easily apply to the multi-party case when the coherent states are generalized in this way.

From this definition, Sugita argues that a separable state is represented by a localized wave packet in phase space. Since coherent states are the most localized states in the Husimi representation, it is argued that delocalization of the Husimi distribution implies correlation - hence entanglement - between the particles. This delocalization can be measured by the R\'enyi-Wehrl entropy, which represents the effective volume occupied by the Husimi distribution.

The majority of Ref. \cite{Sug01} is devoted to constructing explicit formulas for the R\'enyi-Wehrl entropy in terms of the moments. The calculation of the moments is simplified via a group-theoretical construction. When applied to a system of two qubits, it's shown that the concurrence \cite{Woo98} may be expressed in terms of the second moment of the $Q$-function. More strikingly, when applied to a system of three qubits, the expression for the moment contains all three bipartite concurrence terms, as well as the three tangle. In this way this measure captures all classes of entanglement, not only bipartite or otherwise.

Applying the results of Sugita to explain the entanglement behaviour we observe is now straight forward.  Well below the critical parameter, the ground state $Q$-function will be localized around the solitary fixed point, and hence have low degree of entanglement. At the other limit, well above the critical parameter, the ground state consists of a superposition of two, well-separated (and hence almost orthogonal) states. For a bipartite system, this implies that the entanglement of formation will be $1$. In between these two limits the distribution is spread between the three fixed points, with a greater degree of entanglement. As $j$ becomes very large, and the distributions may be more localized, the point at which the ground state distribution is most delocalized will move closer towards the bifurcation point.

In this way, it is the pitchfork structure of the bifurcation that is vital. The loss of stability of the original fixed point, coupled with the emergence of two degenerate stable points, as opposed to a single emergent point, which is the case for other classes of bifurcation, results in the characteristic peak in the ground state entanglement.

We argue that this result should hold for a general bipartite system, and put forth the following conjecture:

\begin{con}[Ground State Entanglement]
Consider a quantum Hamiltonian $\op{H}(g)$ which depends smoothly on a parameter $g$ and which acts on a bipartite Hilbert space $V_1(P) \otimes V_2(P)$. The parameter $P$ allows one to take a well-defined classical limit. Suppose that $\mathcal{H}(g)$ is the well-defined classical limit of $\op{H}(g)$ and that there is a supercritical pitchfork bifurcation of the fixed points at the critical parameter, $g=g_c$. Let the von Neumann entropy $S$, be the measure of entanglement. Then $S(g)$, the entanglement of the ground state of $\op{H}(g)$,  is a maximum with respect to $g$ at $g_{qc}(P)$ where $g_{qc}(P) \rightarrow g_c$ in the classical limit.
\end{con}

\section{Summary}\label{Sec::sum}

We have illustrated how the entanglement in the ground state of a simple coupled tops model can be associated with a bifurcation of the classical fixed points. Following this observation, we have argued why this result should be generalizable to any quantum system with an appropriate classical limit which exhibits a pitchfork bifurcation.

We have already found other instances where our conjecture holds. In particular, for the Dicke model studied  in Ref. \cite{LEB03}. This system exhibits a QPT, which corresponds to the bifurcation. This system is the subject of a future article \cite{MGHM04}.

The classical bifurcation having a signature - namely the entanglement spike - in the quantum regime is not that surprising. Most classical characteristics arise in the quantum regime, although the signature here - entanglement - is uniquely quantum. Here we have a correspondence between the classical and quantum in stationary states of the system. Predominantly, quantum-classical correspondence has been considered with respect to the dynamics, especially in the case of chaotic systems (for example \cite{JE01,Eme01a,FMT03,WGSH04,GASD04}). The dynamical generation of entanglement is argued to be related to the underlying chaos in many-body systems - classical chaos implies a greater degree of entanglement. The simple coupled tops system offers an excellent test bed for further investigation into the relation between chaos and entanglement due to its rich (classical) dynamical structure.

\begin{acknowledgements}
The authors would like to thank Ben Skellett, Tobias Osborne and Ben Toner for valuable discussions and Jacqui Wilton, Michael Bremner and Mohan Sarovar for assistance with the manuscript. This work has been supported by the Australian Research Council.
\end{acknowledgements}

\appendix

\section{Semiclassical Limit}\label{scl}

To take the appropriate semiclassical limit we first consider the correlation function between two operators, $\op{X},\op{Y}$,
\[\label{cov}
\langle\op{X},\op{Y}\rangle = \langle\ox\oy\rangle - \langle\ox\rangle\langle\oy\rangle,
\]
also known as the {\it covariance}. In our case, all operators are elements of the $SU(2)$ group of total angular momentum operators. The scaling of the covariance with respect to the total angular momentum eigenvalue, $\sqrt{j(j+1)}$, is
\[
\langle\op{X},\op{Y}\rangle = \mathcal{O}(\sqrt{j(j+1)}).
\]
i.e., the covariance is of order not exceeding $j$. Conversely, both $\langle\ox\oy\rangle$ and  $\langle\ox\rangle\langle\oy\rangle$ are $\mathcal{O}(j(j+1))$. Re-expressing Eq. \eqrf{cov}, and dividing through by $j(j+1)$ yields
\[\label{exp}
\frac{\langle\ox\oy\rangle}{j(j+1)} = \frac{\langle\ox\rangle}{\sqrt{j(j+1)}}\frac{\langle\oy\rangle}{\sqrt{j(j+1)}} + \mathcal{O}\left(\frac{1}{\sqrt{j(j+1)}}\right).
\]
Taking the limit of $j\rightarrow \infty$, meaning $\sqrt{j(j+1)} \rightarrow j$, gives
\[ \frac{\langle\ox\oy\rangle}{j^2} \approx \frac{\langle\ox\rangle}{j}\frac{\langle\oy\rangle}{j}\].
Thus by defining  variables as the expectation values
\begin{equation}\label{semidef}
L_{a\alpha } = \frac{\langle \op{J}_{a\alpha }\rangle}{j},
\end{equation}
which are simply real numbers, allows the expectation values of products of operators to be factorized in the semiclassical limit. The semiclassical dynamics are then obtained from the Heisenberg operator equations of motion by replacing the operators in the above differential equation with these expectation values. Similarly, the semiclassical Hamiltonian, $E$ is obtained by taking the expectation value of the Hamiltonian, scaling by $1/j$,
\begin{equation}
E = \frac{\langle \op{H}\rangle}{j} = L_{x1} + L_{x2} + \mu L_{z1}L_{z2}.
\end{equation}

\bibliography{ent_bif}

\begin{thebibliography}{40}
\expandafter\ifx\csname natexlab\endcsname\relax\def\natexlab#1{#1}\fi
\expandafter\ifx\csname bibnamefont\endcsname\relax
  \def\bibnamefont#1{#1}\fi
\expandafter\ifx\csname bibfnamefont\endcsname\relax
  \def\bibfnamefont#1{#1}\fi
\expandafter\ifx\csname citenamefont\endcsname\relax
  \def\citenamefont#1{#1}\fi
\expandafter\ifx\csname url\endcsname\relax
  \def\url#1{\texttt{#1}}\fi
\expandafter\ifx\csname urlprefix\endcsname\relax\def\urlprefix{URL }\fi
\providecommand{\bibinfo}[2]{#2}
\providecommand{\eprint}[2][]{\url{#2}}

\bibitem[{\citenamefont{{M.A. Nielsen and I.L. Chuang}}(2000)}]{Nie00}
\bibinfo{author}{\bibnamefont{{M.A. Nielsen and I.L. Chuang}}},
  \emph{\bibinfo{title}{Quantum computation and quantum information}}
  (\bibinfo{publisher}{Cambridge University Press},
  \bibinfo{address}{Cambridge}, \bibinfo{year}{2000}).

\bibitem[{\citenamefont{{S. Sachdev}}(1999)}]{Sac99}
\bibinfo{author}{\bibnamefont{{S. Sachdev}}}, \emph{\bibinfo{title}{Quantum
  phase transitions}} (\bibinfo{publisher}{Cambridge University Press},
  \bibinfo{address}{Cambridge}, \bibinfo{year}{1999}).

\bibitem[{\citenamefont{Osborne and Nielsen}(2002)}]{ON02}
\bibinfo{author}{\bibfnamefont{T.}~\bibnamefont{Osborne}} \bibnamefont{and}
  \bibinfo{author}{\bibfnamefont{M.}~\bibnamefont{Nielsen}},
  \bibinfo{journal}{Phys. Rev. A} \textbf{\bibinfo{volume}{66}},
  \bibinfo{pages}{032110} (\bibinfo{year}{2002}).

\bibitem[{\citenamefont{{A. Osterloh, L. Amico, G. Falci and R.
  Fazio}}(2002)}]{OAF02}
\bibinfo{author}{\bibnamefont{{A. Osterloh, L. Amico, G. Falci and R. Fazio}}},
  \bibinfo{journal}{Nature} \textbf{\bibinfo{volume}{416}},
  \bibinfo{pages}{608} (\bibinfo{year}{2002}).

\bibitem[{\citenamefont{{G. Vidal, J.I. Latorre, E. Rico and A.
  Kitaev}}(2003)}]{VLR02}
\bibinfo{author}{\bibnamefont{{G. Vidal, J.I. Latorre, E. Rico and A.
  Kitaev}}}, \bibinfo{journal}{Phys. Rev. Lett} \textbf{\bibinfo{volume}{90}},
  \bibinfo{pages}{227902} (\bibinfo{year}{2003}).

\bibitem[{\citenamefont{Latorre et~al.}(2004)\citenamefont{Latorre, Rico, and
  Vidal}}]{LRV03}
\bibinfo{author}{\bibfnamefont{J.~I.} \bibnamefont{Latorre}},
  \bibinfo{author}{\bibfnamefont{E.}~\bibnamefont{Rico}}, \bibnamefont{and}
  \bibinfo{author}{\bibfnamefont{G.}~\bibnamefont{Vidal}},
  \bibinfo{journal}{Quant. Inf. and Comp.} \textbf{\bibinfo{volume}{4}},
  \bibinfo{pages}{42} (\bibinfo{year}{2004}).

\bibitem[{\citenamefont{{T. Costi and R.H. McKenzie}}(2003)}]{CM03}
\bibinfo{author}{\bibnamefont{{T. Costi and R.H. McKenzie}}},
  \bibinfo{journal}{Phys. Rev. A} \textbf{\bibinfo{volume}{68}},
  \bibinfo{pages}{034301} (\bibinfo{year}{2003}).

\bibitem[{\citenamefont{{A.P. Hines, R.H. McKenzie and G.J.
  Milburn}}(2003)}]{Hin03}
\bibinfo{author}{\bibnamefont{{A.P. Hines, R.H. McKenzie and G.J. Milburn}}},
  \bibinfo{journal}{Phys. Rev. A} \textbf{\bibinfo{volume}{67}},
  \bibinfo{pages}{013609} (\bibinfo{year}{2003}).

\bibitem[{\citenamefont{Dorner et~al.}(2003)\citenamefont{Dorner, Fedichev,
  Jaksch, Lewenstein, and Zoller}}]{DFJ+02}
\bibinfo{author}{\bibfnamefont{U.}~\bibnamefont{Dorner}},
  \bibinfo{author}{\bibfnamefont{P.}~\bibnamefont{Fedichev}},
  \bibinfo{author}{\bibfnamefont{D.}~\bibnamefont{Jaksch}},
  \bibinfo{author}{\bibfnamefont{M.}~\bibnamefont{Lewenstein}},
  \bibnamefont{and} \bibinfo{author}{\bibfnamefont{P.}~\bibnamefont{Zoller}},
  \bibinfo{journal}{Phys. Rev. Lett.} \textbf{\bibinfo{volume}{91}},
  \bibinfo{pages}{073601} (\bibinfo{year}{2003}).

\bibitem[{\citenamefont{Glendinning}(1994)}]{Gle94}
\bibinfo{author}{\bibfnamefont{P.}~\bibnamefont{Glendinning}},
  \emph{\bibinfo{title}{Stability, instability and chaos: an introduction to
  the theory of nonlinear differential equations}}
  (\bibinfo{publisher}{Cambridge University Press}, \bibinfo{year}{1994}).

\bibitem[{\citenamefont{{S.Schneider and G.J.Milburn}}(2002)}]{Sch01}
\bibinfo{author}{\bibnamefont{{S.Schneider and G.J.Milburn}}},
  \bibinfo{journal}{Phys. Rev. A} \textbf{\bibinfo{volume}{65}},
  \bibinfo{pages}{042107} (\bibinfo{year}{2002}).

\bibitem[{\citenamefont{{J. Tejada, E.M Chudnovsky, E. del Barco, J.M.
  Hernandez and T.P. Spiller}}(2000)}]{Tej00}
\bibinfo{author}{\bibnamefont{{J. Tejada, E.M Chudnovsky, E. del Barco, J.M.
  Hernandez and T.P. Spiller}}}, \bibinfo{type}{Tech. Rep.}
  \bibinfo{number}{HPL-2000-87}, \bibinfo{institution}{Hewlett Packard
  Laboratories} (\bibinfo{year}{2000}).

\bibitem[{\citenamefont{{B.Skellett and C.A.Holmes}}(2002)}]{axe02}
\bibinfo{author}{\bibnamefont{{B.Skellett and C.A.Holmes}}},
  \bibinfo{journal}{Interjournal Complex Systems} \textbf{\bibinfo{volume}{51}}
  (\bibinfo{year}{2002}), \bibinfo{note}{www.interjournal.org}.

\bibitem[{\citenamefont{{D.Gunlycke, S.Bose, V.M.Kendon and
  V.Vedral}}(2001)}]{Gun01}
\bibinfo{author}{\bibnamefont{{D.Gunlycke, S.Bose, V.M.Kendon and V.Vedral}}},
  \bibinfo{journal}{Phys. Rev. A} \textbf{\bibinfo{volume}{64}},
  \bibinfo{pages}{042302} (\bibinfo{year}{2001}).

\bibitem[{\citenamefont{Preskill}()}]{pres}
\bibinfo{author}{\bibfnamefont{J.}~\bibnamefont{Preskill}},
  \bibinfo{howpublished}{http://www.theory.caltech.edu/people
  /preskill/ph229/$\#$lecture (Chap. 5)}.

\bibitem[{\citenamefont{Husimi}(1940)}]{Hus40}
\bibinfo{author}{\bibfnamefont{K.}~\bibnamefont{Husimi}},
  \bibinfo{journal}{Proc. Phys. Math. Soc. Jpn.} \textbf{\bibinfo{volume}{22}},
  \bibinfo{pages}{264} (\bibinfo{year}{1940}).

\bibitem[{\citenamefont{{M. Latka, P. Grigolini and B.J. West}}(1993)}]{Lat93}
\bibinfo{author}{\bibnamefont{{M. Latka, P. Grigolini and B.J. West}}},
  \bibinfo{journal}{Phys. Rev. A} \textbf{\bibinfo{volume}{47}},
  \bibinfo{pages}{6} (\bibinfo{year}{1993}).

\bibitem[{\citenamefont{{G.J. Milburn}}(1986)}]{Mil86}
\bibinfo{author}{\bibnamefont{{G.J. Milburn}}}, \bibinfo{journal}{Phys. Rev. A}
  \textbf{\bibinfo{volume}{33}}, \bibinfo{pages}{1} (\bibinfo{year}{1986}).

\bibitem[{\citenamefont{{B.C. Sanders}}(1989)}]{San89}
\bibinfo{author}{\bibnamefont{{B.C. Sanders}}}, \bibinfo{journal}{Phys. Rev. A}
  \textbf{\bibinfo{volume}{40}}, \bibinfo{pages}{5} (\bibinfo{year}{1989}).

\bibitem[{\citenamefont{{W.A. Lin and L.E. Ballentine}}(1990)}]{Lin90}
\bibinfo{author}{\bibnamefont{{W.A. Lin and L.E. Ballentine}}},
  \bibinfo{journal}{Phys. Rev. Lett.} \textbf{\bibinfo{volume}{65}},
  \bibinfo{pages}{24} (\bibinfo{year}{1990}).

\bibitem[{\citenamefont{{S.B. Lee and M.D. Feit}}(1993)}]{Lee93}
\bibinfo{author}{\bibnamefont{{S.B. Lee and M.D. Feit}}},
  \bibinfo{journal}{Phys. Rev. E} \textbf{\bibinfo{volume}{47}},
  \bibinfo{pages}{6} (\bibinfo{year}{1993}).

\bibitem[{\citenamefont{{G.J. Milburn, R. Laflamme, B.C. Sanders and E.
  Knill}}(2002)}]{MLSK01}
\bibinfo{author}{\bibnamefont{{G.J. Milburn, R. Laflamme, B.C. Sanders and E.
  Knill}}}, \bibinfo{journal}{Phys. Rev. A} \textbf{\bibinfo{volume}{65}},
  \bibinfo{pages}{032316} (\bibinfo{year}{2002}).

\bibitem[{\citenamefont{Appleby}(2000)}]{App00}
\bibinfo{author}{\bibfnamefont{D.}~\bibnamefont{Appleby}},
  \bibinfo{journal}{Int. J. Theor. Phys.} \textbf{\bibinfo{volume}{39}},
  \bibinfo{pages}{2231} (\bibinfo{year}{2000}).

\bibitem[{\citenamefont{{A. Perelomov}}(1986)}]{Per86}
\bibinfo{author}{\bibnamefont{{A. Perelomov}}},
  \emph{\bibinfo{title}{Generalized {C}oherent {S}tates and {T}heir
  {A}pplications}} (\bibinfo{publisher}{Springer-Verlag},
  \bibinfo{address}{Berlin}, \bibinfo{year}{1986}).

\bibitem[{\citenamefont{Sugita}(2003)}]{Sug01}
\bibinfo{author}{\bibfnamefont{A.}~\bibnamefont{Sugita}},
  \bibinfo{journal}{J.Phys. A: Math. Gen.} \textbf{\bibinfo{volume}{36}},
  \bibinfo{pages}{9081} (\bibinfo{year}{2003}).

\bibitem[{\citenamefont{Bohigas et~al.}(1993)\citenamefont{Bohigas, Tomsovic,
  and Ullmo}}]{BTU93}
\bibinfo{author}{\bibfnamefont{O.}~\bibnamefont{Bohigas}},
  \bibinfo{author}{\bibfnamefont{S.}~\bibnamefont{Tomsovic}}, \bibnamefont{and}
  \bibinfo{author}{\bibfnamefont{D.}~\bibnamefont{Ullmo}},
  \bibinfo{journal}{Phys. Rep.} \textbf{\bibinfo{volume}{223}},
  \bibinfo{pages}{43} (\bibinfo{year}{1993}).

\bibitem[{\citenamefont{Berry}(1977)}]{Ber77}
\bibinfo{author}{\bibfnamefont{M.~V.} \bibnamefont{Berry}},
  \bibinfo{journal}{Philos. Trans. R. Soc. London}
  \textbf{\bibinfo{volume}{287}}, \bibinfo{pages}{237} (\bibinfo{year}{1977}).

\bibitem[{\citenamefont{Ketzmerik et~al.}(2000)\citenamefont{Ketzmerik,
  Hufnagel, Steinbach, and Weiss}}]{KHSW00}
\bibinfo{author}{\bibfnamefont{R.}~\bibnamefont{Ketzmerik}},
  \bibinfo{author}{\bibfnamefont{L.}~\bibnamefont{Hufnagel}},
  \bibinfo{author}{\bibfnamefont{F.}~\bibnamefont{Steinbach}},
  \bibnamefont{and} \bibinfo{author}{\bibfnamefont{M.}~\bibnamefont{Weiss}},
  \bibinfo{journal}{Phys Rev. Lett.} \textbf{\bibinfo{volume}{85}},
  \bibinfo{pages}{1214} (\bibinfo{year}{2000}).

\bibitem[{\citenamefont{{F. Haake}}(1991)}]{Haa91}
\bibinfo{author}{\bibnamefont{{F. Haake}}}, \emph{\bibinfo{title}{Quantum
  {S}ignatures of {C}haos}} (\bibinfo{publisher}{Springer, Berlin},
  \bibinfo{year}{1991}).

\bibitem[{\citenamefont{Percival}(1973)}]{Per73}
\bibinfo{author}{\bibfnamefont{I.~C.} \bibnamefont{Percival}},
  \bibinfo{journal}{J, Phys. B} \textbf{\bibinfo{volume}{6}},
  \bibinfo{pages}{L229} (\bibinfo{year}{1973}).

\bibitem[{\citenamefont{Shnirelman}(1974)}]{Shn74}
\bibinfo{author}{\bibfnamefont{A.~I.} \bibnamefont{Shnirelman}},
  \bibinfo{journal}{Usp. Math. Nauk.} \textbf{\bibinfo{volume}{29}},
  \bibinfo{pages}{181} (\bibinfo{year}{1974}).

\bibitem[{\citenamefont{Zheng et~al.}(1990)\citenamefont{Zheng, Feng, and
  Gilmore}}]{ZFG90}
\bibinfo{author}{\bibfnamefont{W.-M.} \bibnamefont{Zheng}},
  \bibinfo{author}{\bibfnamefont{D.~H.} \bibnamefont{Feng}}, \bibnamefont{and}
  \bibinfo{author}{\bibfnamefont{R.}~\bibnamefont{Gilmore}},
  \bibinfo{journal}{Rev. Mod. Phys.} \textbf{\bibinfo{volume}{62}},
  \bibinfo{pages}{867} (\bibinfo{year}{1990}).

\bibitem[{\citenamefont{{W.K. Wooters}}(1998)}]{Woo98}
\bibinfo{author}{\bibnamefont{{W.K. Wooters}}}, \bibinfo{journal}{Phys. Rev.
  Lett.} \textbf{\bibinfo{volume}{80}}, \bibinfo{pages}{2245}
  (\bibinfo{year}{1998}).

\bibitem[{\citenamefont{{N. Lambert, C. Emary and T. Brandes}}(2004)}]{LEB03}
\bibinfo{author}{\bibnamefont{{N. Lambert, C. Emary and T. Brandes}}},
  \bibinfo{journal}{Phys. Rev. Lett.} \textbf{\bibinfo{volume}{92}},
  \bibinfo{pages}{073602} (\bibinfo{year}{2004}).

\bibitem[{\citenamefont{Morrison et~al.}(2004)\citenamefont{Morrison,
  Gilchrist, Hines, and Milburn}}]{MGHM04}
\bibinfo{author}{\bibfnamefont{S.}~\bibnamefont{Morrison}},
  \bibinfo{author}{\bibfnamefont{A.}~\bibnamefont{Gilchrist}},
  \bibinfo{author}{\bibfnamefont{A.~P.} \bibnamefont{Hines}}, \bibnamefont{and}
  \bibinfo{author}{\bibfnamefont{G.~J.} \bibnamefont{Milburn}}
  (\bibinfo{year}{2004}), \bibinfo{note}{{\it in preparation}}.

\bibitem[{\citenamefont{{J.V. Emerson}}(2001)}]{JE01}
\bibinfo{author}{\bibnamefont{{J.V. Emerson}}}, Ph.D. thesis,
  \bibinfo{school}{Simon Fraser University} (\bibinfo{year}{2001}).

\bibitem[{\citenamefont{{J. Emerson and L.E. Ballentine}}(2001)}]{Eme01a}
\bibinfo{author}{\bibnamefont{{J. Emerson and L.E. Ballentine}}},
  \bibinfo{journal}{Phys. Rev. A} \textbf{\bibinfo{volume}{63}},
  \bibinfo{pages}{052103} (\bibinfo{year}{2001}).

\bibitem[{\citenamefont{Fujisaki et~al.}(2003)\citenamefont{Fujisaki, Miyadera,
  and Tanaka}}]{FMT03}
\bibinfo{author}{\bibfnamefont{H.}~\bibnamefont{Fujisaki}},
  \bibinfo{author}{\bibfnamefont{T.}~\bibnamefont{Miyadera}}, \bibnamefont{and}
  \bibinfo{author}{\bibfnamefont{A.}~\bibnamefont{Tanaka}},
  \bibinfo{journal}{Phys. Rev. E} \textbf{\bibinfo{volume}{67}},
  \bibinfo{pages}{066201} (\bibinfo{year}{2003}).

\bibitem[{\citenamefont{Wang et~al.}(2004)\citenamefont{Wang, Ghose, Sanders,
  and Hu}}]{WGSH04}
\bibinfo{author}{\bibfnamefont{X.}~\bibnamefont{Wang}},
  \bibinfo{author}{\bibfnamefont{S.}~\bibnamefont{Ghose}},
  \bibinfo{author}{\bibfnamefont{B.~C.} \bibnamefont{Sanders}},
  \bibnamefont{and} \bibinfo{author}{\bibfnamefont{B.}~\bibnamefont{Hu}},
  \bibinfo{journal}{Phys. Rev. E} \textbf{\bibinfo{volume}{70}},
  \bibinfo{pages}{016217} (\bibinfo{year}{2004}).

\bibitem[{\citenamefont{Ghose et~al.}()\citenamefont{Ghose, Alsing, Sanders,
  and Deutsch}}]{GASD04}
\bibinfo{author}{\bibfnamefont{S.}~\bibnamefont{Ghose}},
  \bibinfo{author}{\bibfnamefont{P.}~\bibnamefont{Alsing}},
  \bibinfo{author}{\bibfnamefont{B.~C.} \bibnamefont{Sanders}},
  \bibnamefont{and} \bibinfo{author}{\bibfnamefont{I.~H.}
  \bibnamefont{Deutsch}}, \eprint{quant-ph/0409133}.

\end{thebibliography}
\end{document}